\documentclass[10pt]{amsart}%
\usepackage{amsbsy,amssymb,amscd,amsfonts,latexsym,amstext,delarray,
amsmath,epsfig,graphicx,bm, url}
\usepackage{hyperref}
\usepackage{amsmath}
\usepackage{amsfonts}
\usepackage{amssymb}
\usepackage{graphicx}%
\input xypic
\newtheorem{thm}{Theorem}[section]
\newtheorem{prop}[thm]{Proposition}
\newtheorem{cor}[thm]{Corollary}
\newtheorem{lem}[thm]{Lemma}

\numberwithin{equation}{section}

\def\text{\hbox}

\begin{document}
\title{Inner fluctuations of the spectral action}
\author[Connes]{Alain Connes}
\author[Chamseddine]{Ali H. Chamseddine}
\address{A.~Connes: Coll\`ege de France \\
3, rue d'Ulm \\
Paris, F-75005 France\\
I.H.E.S. and Vanderbilt University\\
A.~Chamseddine: Physics Department, American University of Beirut, Lebanon}
\email{chams@aub.edu.lb}
\maketitle

\begin{abstract}
We prove in the general framework of noncommutative geometry that the inner
fluctuations of the spectral action can be computed as residues and give
exactly the counterterms for the Feynman graphs with fermionic internal lines.
We show that for geometries of dimension less or equal to four the obtained
terms add up to a sum of a Yang-Mills action with a Chern-Simons action.

\end{abstract}
\tableofcontents

\begin{verse}
\textit{We dedicate this paper to Daniel Kastler on his eightieth birthday}
\end{verse}

\section{Introduction}

The spectral action is defined as a functional on noncommutative geometries.
Such a geometry is specified by a fairly simple data of operator theoretic
nature, namely a spectral triple
\begin{equation}
({\mathcal{A}},{\mathcal{H}},D), \label{spectrip}%
\end{equation}
where ${\mathcal{A}}$ is a noncommutative algebra with involution $\ast$,
acting in the Hilbert space ${\mathcal{H}}$ while $D$ is a self-adjoint
operator with compact resolvent and such that,
\begin{equation}
\lbrack D,a]\ \hbox{is bounded}\ \forall\,a\in{\mathcal{A}}\,.
\label{boundedcom}%
\end{equation}
Additional structures such as the ${\mathbb{Z}}/2{\mathbb{Z}}$ grading
$\gamma$ in the even case and the real structure $J$ of ${\mathcal{H}}$ will
play little role below, but can easily be taken into account.

The spectral action fulfills two basic properties

\begin{itemize}
\item It only depends upon the spectrum of $D$.

\item It is additive for direct sums of noncommutative geometries.
\end{itemize}

It is given in general by the expression
\begin{equation}
\mathrm{Trace}\,(f(D/\Lambda)), \label{specact}%
\end{equation}
where $f$ is a positive even function of the real variable and the parameter
$\Lambda$ fixes the mass scale. The dimension of a noncommutative geometry is
not a number but a spectrum, the \emph{dimension spectrum} (\textit{cf.\/}%
\ \cite{[C-M2]}) which is the subset $\Pi$ of the complex plane ${\mathbb{C}}$
at which the spectral functions have singularities. Under the hypothesis that
the dimension spectrum is \emph{simple} \textit{i.e.\/}\ that the spectral
functions have at most simple poles, the residue at the pole defines a far
reaching extension (\textit{cf.\/}\ \cite{[C-M2]}) of the fundamental integral
in noncommutative geometry given by the Dixmier trace (\textit{cf.\/}%
\ \cite{book}). This extends to the framework of spectral triples the Wodzicki
residue (originally defined for pseudodifferential operators on standard
manifolds) as a trace on the algebra of operators generated by ${\mathcal{A}}$
and powers of $D$ so that
\begin{equation}
P\rightarrow{\int\!\!\!\!\!\!-}\,P\in{\mathbb{C}}\,,\quad{\int\!\!\!\!\!\!-}%
\,P_{1}\,P_{2}=\,{\int\!\!\!\!\!\!-}\,P_{2}\,P_{1}. \label{residue}%
\end{equation}

Both this algebra and the functional \eqref{residue} do not depend on the
detailed knowledge of the \emph{metric} defined by $D$ and the residue is
unaltered by a change $D\rightarrow D^{\prime}$ of $D$ such that the
difference
\[
\mathrm{Log}\,D^{\prime}-\,\mathrm{Log}\,D,
\]
is a \emph{bounded} operator with suitable regularity. In other words the
residue only depends on the quasi-isometry class of the noncommutative metric.

In this generality the spectral action \eqref{specact} can be expanded in
decreasing powers of the scale $\Lambda$ in the form
\begin{equation}
\mathrm{Trace}\,(f(D/\Lambda))\sim\,\sum_{k\in\,\Pi^{+}}\,f_{k}\,\Lambda
^{k}\,{\int\!\!\!\!\!\!-}\,|D|^{-k}\,+\,f(0)\,\zeta_{D}%
(0)+\,o(1),\label{expansion}%
\end{equation}
where $\Pi^{+}$ is the positive part of the dimension spectrum $\Pi$. The
function $f$ only appears through the scalars
\begin{equation}
f_{k}=\,\,\int_{0}^{\infty}f(v)\,v^{k-1}\,dv.\label{coeff}%
\end{equation}
One lets
\begin{equation}
\zeta_{D}(s)=\,\mathrm{Tr}\,(|D|^{-s}),\label{zeta}%
\end{equation}
and regularity at $s=0$ is assumed.

\begin{figure}[ptb]
\begin{center}
\includegraphics[scale=0.7]{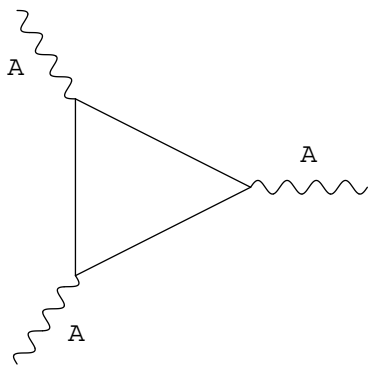}
\end{center}
\caption{The triangle graph. }%
\label{Fig3legs}%
\end{figure}

\medskip Both the gauge bosons and the Feynman graphs with fermionic internal
lines can be readily defined in the above generality of a noncommutative
geometry $({\mathcal{A}},{\mathcal{H}},D)$ (\textit{cf.\/}\ \cite{essay}).
Indeed, as briefly recalled at the beginning of section \ref{fluct}, the inner
fluctuations of the metric coming from the Morita equivalence ${\mathcal{A}%
}\sim{\mathcal{A}}$ generate perturbations of $D$ of the form $D\rightarrow
D^{\prime}=D+A$ where the $A$ plays the role of the gauge potentials and is a
self-adjoint element of the bimodule
\begin{equation}
\Omega_{D}^{1}=\,\{\;\sum\,a_{j}\,[D,b_{j}]\;;\,a_{j},\,b_{j}\in{\mathcal{A}%
}\}. \label{bim}%
\end{equation}
The line element $ds=\,D^{-1}$ plays the role of the Fermion propagator so
that the value $U(\Gamma_{n})$ of one loop graphs $\Gamma_{n}$ with fermionic
internal lines and $n$ external bosonic lines (such as the triangle graph of
Figure \ref{Fig3legs}) is easy to obtain and given at the formal level by,
\[
U(\Gamma_{n})=\,\mathrm{Tr}((AD^{-1})^{n}).
\]
These graphs diverge in dimension $4$ for $n\leq4$ and the residue at the pole
in dimensional regularization can be computed and expressed as
\[
{\int\!\!\!\!\!\!-}(AD^{-1})^{n},
\]
as will be shown in \cite{CManom}.

\medskip In this paper we analyze how the spectral action behaves under the
inner fluctuations. The main results are

\begin{figure}[ptb]
\begin{center}
\includegraphics[scale=0.7]{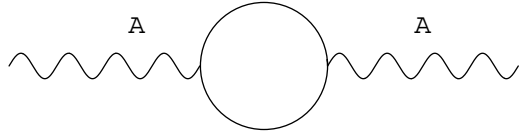}
\end{center}
\caption{The self-energy graph. }%
\label{Fig2legs}%
\end{figure}

\begin{figure}[ptb]
\begin{center}
\includegraphics[scale=0.7]{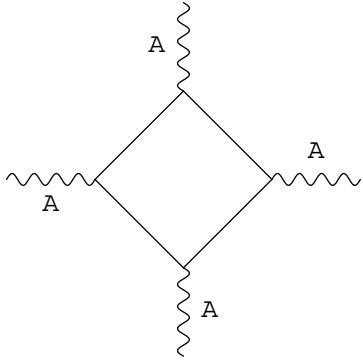}
\end{center}
\caption{The quartic graph. }%
\label{Fig4legs}%
\end{figure}

\begin{itemize}
\item In dimension $4$ the variation of the spectral action under inner
fluctuations gives the local counterterms for the fermionic graphs of Figures
\ref{Fig1leg}, \ref{Fig2legs}, \ref{Fig3legs} and \ref{Fig4legs} respectively
\[
\zeta_{D+A}(0)-\zeta_{D}(0)=\,-\,{\int\!\!\!\!\!\!-}AD^{-1}+\frac{1}{2}%
\,{\int\!\!\!\!\!\!-}\,(AD^{-1})^{2}-\frac{1}{3}\,{\int\!\!\!\!\!\!-}%
\,(AD^{-1})^{3}+\frac{1}{4}\,{\int\!\!\!\!\!\!-}\,(AD^{-1})^{4},
\]

\item Assuming that the tadpole graph of Figure \ref{Fig1leg} vanishes the
above variation is the sum of a Yang-Mills action and a Chern-Simons action
relative to a cyclic $3$-cocycle on ${\mathcal{A}}$.
\end{itemize}

As a corollary, combining both results we obtain that the variation under
inner fluctuations of the scale independent terms of the spectral action is
given (\textit{cf.\/}\ Theorem \ref{main} for precise notations) in dimension
$4$ by
\begin{equation}
\zeta_{D+A}(0)-\zeta_{D}(0)=\,\,\frac{1}{4}\,\int_{\tau_{0}}\,(dA+\,A^{2}%
)^{2}-\frac{1}{2}\,\int_{\psi}\,(AdA+\,\frac{2}{3}A^{3}). \label{chern+ym}%
\end{equation}

The conceptual meaning of the above tadpole condition is that the original
noncommutative geometry $({\mathcal{A}},{\mathcal{H}},D)$ is a critical point
for the ($\Lambda$-independent part of the) spectral action, which is a
natural hypothesis. The functional $\tau_{0}$ is a Hochschild $4$-cocycle but
in general not a cyclic cocycle. In particular, as explained in details in
\cite{book} Chapter VI, the expression
\begin{equation}
\int_{\tau_{0}}\,(dA+\,A^{2})^{2}, \label{ym}%
\end{equation}
coincides with the Yang-Mills action functional provided that $\tau_{0}\geq0$
\textit{i.e.\/}\ that $\tau_{0}$ is a positive Hochschild cocycle. The
Hochschild cocycle $\tau_{0}$ cannot be cyclic unless the expression
\eqref{ym} vanishes.

We show at the end of the paper that the cyclic cohomology class of the cyclic
three cocycle $\psi$ is determined modulo the image of the boundary operator
$B$ and that the pairing of $\psi$ with the $K_{1}$-group is trivial. This
shows that under rather general assumptions one can eliminate $\psi$ by a
suitable redefinition of $\tau_{0}$ (see Proposition \ref{ambig}).

The meaning of the vanishing of $\psi$ together with positivity of $\tau_{0}$
is that the original noncommutative geometry $({\mathcal{A}},{\mathcal{H}},D)$
is at a \emph{stable} critical point as far as the inner fluctuations are
concerned. In fact it gives in that case an absolute minimum for the (scale
independent terms of the) spectral action in the corresponding class modulo
inner fluctuations. We end the paper with the corresponding open questions :
elimination of $\psi$ and positivity of the $4$-cocycle $\tau_{0}$.

\bigskip

\section{Inner fluctuations of the metric and the spectral action}

\label{fluct}

The inner fluctuations of the noncommutative metric appear through the simple
issue of Morita equivalence. Indeed let ${\mathcal{B}}$ be the algebra of
endomorphisms of a finite projective (right) module ${\mathcal{E}}$ over
${\mathcal{A}}$
\begin{equation}
{\mathcal{B}}=\mathrm{End}_{{\mathcal{A}}}({\mathcal{E}}). \label{endE}%
\end{equation}
Given a spectral triple $({\mathcal{A}},{\mathcal{H}},D)$ one easily gets a
representation of ${\mathcal{B}}$ in the Hilbert space
\[
{\mathcal{H}}^{\prime}=\,{\mathcal{E}}\otimes_{{\mathcal{A}}}\,{\mathcal{H}%
}\,.
\]
But to define the analogue $D^{\prime}$ of the operator $D$ for ${\mathcal{B}%
}$ requires the choice of a \textit{hermitian connection} on ${\mathcal{E}}$.
Such a connection $\nabla$ is a linear map $\nabla:{\mathcal{E}}%
\rightarrow{\mathcal{E}}\otimes_{{\mathcal{A}}}\Omega_{D}^{1}$ satisfying the
following rules (\cite{book})
\begin{equation}
\nabla(\xi a)=(\nabla\xi)a+\xi\otimes da{\,,\quad\forall\,}\,\xi
\in{\mathcal{E}}\ ,\ a\in{\mathcal{A,}} \label{connect}%
\end{equation}%
\begin{equation}
(\xi,\nabla\eta)-(\nabla\xi,\eta)=d(\xi,\eta){\,,\quad\forall\,}\xi,\eta
\in{\mathcal{E,}} \label{connect1}%
\end{equation}
where $da=[D,a]$ and where $\Omega_{D}^{1}\subset{\mathcal{L}}({\mathcal{H}})$
is the ${\mathcal{A}}$--bimodule \eqref{bim}. The operator $D^{\prime}$ is
then given by
\begin{equation}
D^{\prime}(\xi\otimes\eta)=\,\xi\otimes\,D\,\eta+\,\nabla(\xi)\eta\,.
\label{connect1}%
\end{equation}

\smallskip

Any algebra ${\mathcal{A}}$ is Morita equivalent to itself and when one
applies the above construction with ${\mathcal{E}}={\mathcal{A}}$ one gets the
inner deformations of the spectral geometry. These replace the operator $D$
by
\begin{equation}
D\rightarrow D+A,
\end{equation}
where $A=A^{\ast}$ is an arbitrary selfadjoint element of $\Omega_{D}^{1}$
where we disregard the real structure for simplicity. To incorporate the real
structure one replaces the algebra ${\mathcal{A}}$ by its tensor product
${\mathcal{A}}\otimes{\mathcal{A}}^{o}$ with the opposite algebra.

\subsection{Pseudodifferential calculus}

\noindent\newline As developed in \cite{[C-M2]} one has under suitable
regularity hypothesis on the spectral geometry $({\mathcal{A}},{\mathcal{H}%
},D)$ an analogue of the pseudodifferential calculus. We briefly recall the
main ingredients here. We say that an operator $T$ in ${\mathcal{H}}$ is
\textit{smooth} iff
\begin{equation}
t\rightarrow F_{t}(T)=e^{it|D|}\,T\,e^{-it|D|}\in C^{\infty}({\mathbb{R}%
},{\mathcal{L}}({\mathcal{H}})), \label{flow}%
\end{equation}
and let $OP^{0}$ be the algebra of smooth operators. Any smooth operator $T$
belongs to the domains of $\delta^{n}$, where the derivation $\delta$ is
defined by
\begin{equation}
\delta(T)=|D|\,T-T\,|D|=[|D|,T]. \label{delta}%
\end{equation}
The analogue of the Sobolev spaces are given by
\[
{\mathcal{H}}_{s}=\mathrm{Dom}\,|D|^{s}\qquad s\geq0\,,\quad{\mathcal{H}}%
_{-s}=({\mathcal{H}}_{s})^{\ast}\,,\;s<0\,.
\]
For any smooth operator $T$ one has (\textit{cf.\/}\ \cite{[C-M2]})
$T\,{\mathcal{H}}_{s}\subset{\mathcal{H}}_{s}\,$ and we let
\[
OP^{\alpha}=\{T\,;\ |D|^{-\alpha}\,T\in OP^{0}\}\,.
\]
We work in dimension $\leq4$ which means that $ds=\,D^{-1}$ is an
infinitesimal of order $\frac{1}{4}$ and thus that for $N>4$, $OP^{-N}$ is
inside trace class operators. In general we work modulo operators of large
negative order, \textit{i.e.\/}\ $\mathrm{mod}\,OP^{-N}$ for large $N$. We let
${\mathcal{D}}({\mathcal{A}})$ be the algebra generated by ${\mathcal{A}}$ and
$D$ considered first at the formal level. The main point is the following
lemma \cite{[C-M2]} which allows to multiply together
\emph{pseudodifferential} operators of the form
\begin{equation}
P\,D^{-2n}\,,\qquad P\in{\mathcal{D}}({\mathcal{A}}). \label{psido}%
\end{equation}
One lets $\nabla(T)=D^{2}\,T-T\,D^{2}$.

\begin{lem}
\label{newton0}\cite{[C-M2]} Let $T \in OP^{0}$.

\begin{itemize}
\item[\textrm{a)}] $\nabla^{n}(T)\in OP^{n}\qquad\forall\,n\in{\mathbb{N}}$

\item[\textrm{b)}] $D^{-2} \, T = \ \sum_{0}^{n} (-1)^{k} \, \nabla^{k} (T) \,
D^{-2k-2} + (-1)^{n+1} \, D^{-2} \, \nabla^{n+1} (T) \, D^{-2n-2}$.

\item[\textrm{c)}] The remainder $R_{n}=D^{-2} \, \nabla^{n+1} (T)\,
D^{-2n-2}$ belongs to $OP^{-(n+3)}$.
\end{itemize}
\end{lem}

\proof a) The equality
\begin{equation}
|D|T|D|^{-1}=T+\beta(T)\,,\quad\beta(T)=\delta(T)\,|D|^{-1},
\end{equation}
shows that for $T\in OP^{0}$ one has
\begin{equation}
D^{2}\,T\,D^{-2}=T+2\,\beta\,(T)+\beta^{2}(T)\in OP^{0}. \label{stable1}%
\end{equation}
Similarly one has,
\begin{equation}
D^{-2}\,T\,D^{2}\in OP^{0}\,. \label{stable2}%
\end{equation}
This shows that in the definition of $OP^{\alpha}$ one can put $|D|^{-\alpha}$
on either side.

To prove a) we just need to check that $\nabla(T)\in OP^{1}$ and then proceed
by induction. We have $\nabla(T)=D^{2}\,T-T\,D^{2}=(D^{2}\,T\,D^{-2}%
-T)\,D^{2}=(2\beta\,(T)+\beta^{2}(T))\,D^{2}=2\delta\,(T)\,|D|+\delta^{2}%
(T)$,
\begin{equation}
\nabla(T)=2\delta\,(T)\,|D|+\delta^{2}(T), \label{eq1.16}%
\end{equation}
which belongs to $OP^{1}$.

\smallskip

b) For $n=0$ the statement follows from
\begin{equation}
D^{-2}\,T=T\,D^{-2}-D^{-2}\,\nabla(T)\,D^{-2}. \label{oneover}%
\end{equation}
Next assume we proved the result for $(n-1)$. To get it for $n$ we must show
that
\begin{align}
&  (-1)^{n}\,\nabla^{n}(T)\,D^{-2n-2}+(-1)^{n+1}\,D^{-2}\,\nabla
^{n+1}(T)\,D^{-2n-2}\nonumber\\
&  =(-1)^{n}\,D^{-2}\,\nabla^{n}(T)\,D^{-2n}\,. \label{eq1.17}%
\end{align}
Multiplying by $D^{2n}$ on the right, with $T^{\prime}=(-1)^{n}\,\nabla
^{n}(T)$, we need to show that
\[
T^{\prime}\,D^{-2}-D^{-2}\,\nabla(T^{\prime})\,D^{-2}=D^{-2}\,T^{\prime},
\]
which is (\ref{oneover}).

c) Follows from a).\endproof

Thus when working $\mathrm{mod}\,OP^{-N}$ for large $N$ one can write
\begin{equation}
D^{-2}\,T\sim\ \sum_{0}^{\infty}(-1)^{k}\,\nabla^{k}(T)\,D^{-2k-2},
\label{newton}%
\end{equation}
and this allows to compute the product in the algebra $\Psi{\mathcal{D}}$ of
operators which, $\mathrm{modulo}\,OP^{-N}$ for any $N,$ are of the form
\eqref{psido}. Such operators will be called pseudodifferential.

\subsection{The operator $\mathrm{Log}\,(D+A)^{2}-\mathrm{Log}\,(D^{2})$}

\noindent\newline

We let $A$ be a gauge potential,
\begin{equation}
A=\sum a_{i}\,[D,b_{i}]\,;\ a_{i},b_{i}\in{\mathcal{A}}\,,\ \quad A=A^{\ast},
\end{equation}
and we consider the operator $X$ defined from the square of the self-adjoint
operator $D+A$,
\begin{equation}
(D+A)^{2}=\,D^{2}+\,X\,,\quad X=AD+DA+A^{2}\,.
\end{equation}
The following lemma is an adaptation to our set-up of a classical result in
the pseudodifferential calculus on manifolds, \medskip

\begin{lem}
\label{difflog}
\[
Y = \mathrm{Log} \, (D+A)^{2} - \mathrm{Log} \, (D^{2}) \in\Psi{\mathcal{D}}
\cap\,OP^{-1}\, .
\]

\end{lem}

\proof We start with the equality $(a>0)$
\begin{equation}
\mathrm{Log}\,a=\int_{0}^{\infty}\left(  \frac{1}{\lambda+1}-\frac{1}%
{\lambda+a}\right)  \,d\lambda,
\end{equation}
and apply it to both $D^{2}$ and $(D+A)^{2}=D^{2}+X$ to get,
\begin{equation}
Y=\int_{0}^{\infty}\left(  \frac{1}{\lambda+D^{2}}-\frac{1}{\lambda+D^{2}%
+X}\right)  \,d\lambda\,. \label{logy}%
\end{equation}
One has
\[
(\lambda+D^{2}+X)^{-1}=((1+X\,(D^{2}+\lambda)^{-1})(D^{2}+\lambda
))^{-1}=(D^{2}+\lambda)^{-1}(1+X(D^{2}+\lambda)^{-1})^{-1},
\]
and one can expand,
\begin{equation}
(1+X\,(D^{2}+\lambda)^{-1})^{-1}=\sum_{0}^{\infty}\,(-1)^{n}\,(X(D^{2}%
+\lambda)^{-1})^{n}\,. \label{invX}%
\end{equation}
In this expansion the remainder is, up to sign,
\begin{equation}
(X(D^{2}+\lambda)^{-1})^{n+1}\,(1+X(D^{2}+\lambda)^{-1})^{-1}=R_{n}%
(\lambda)\,. \label{Rn}%
\end{equation}
Here $X\in OP^{1}$ by construction so that a rough estimate of the order of
the remainder is given by
\begin{equation}
\int X^{n+1}(D^{2}+\lambda)^{-(n+1)}\,d\lambda\sim X^{n+1}(D^{2})^{-n}%
\sim|D|^{n+1-2n}\,. \label{simpleint}%
\end{equation}
Now in lemma~\ref{newton}~b) we can use $D^{2}+\lambda$ instead of $D^{2}$.
This does not alter $\nabla$ since
\begin{equation}
\lbrack D^{2}+\lambda\,,\,T]=[D^{2},T]\,, \label{indep}%
\end{equation}
and we thus get,
\begin{align}
(D^{2}+\lambda)^{-1}\,T  &  =\sum_{0}^{n}\,(-1)^{k}\,\nabla^{k}(T)(D^{2}%
+\lambda)^{-(k+1)}\nonumber\\
&  +(-1)^{n+1}\,(D^{2}+\lambda)^{-1}\,\nabla^{n+1}(T)(D^{2}+\lambda
)^{-(n+1)}\,. \label{moveright}%
\end{align}
Thus using (\ref{invX}) the integrand in (\ref{logy}) is up to a remainder,
\begin{align}
&  (D^{2}+\lambda)^{-1}\,X(D^{2}+\lambda)^{-1}-(D^{2}+\lambda)^{-1}%
\,X(D^{2}+\lambda)^{-1}\,X(D^{2}+\lambda)^{-1}\nonumber\\
&  +\cdots+(-1)^{k+1}((D^{2}+\lambda)^{-1}\,X)^{k}(D^{2}+\lambda)^{-1}%
+\cdots\label{integrand}%
\end{align}
Using (\ref{moveright}) one can move all the $(D^{2}+\lambda)^{-1}$ to the
right at the expense of replacing $X$'s by $\nabla^{k_{j}}(X)$ and increasing
the $n$ in $(D^{2}+\lambda)^{-n}$. Thus using, ($n\geq2$)
\begin{equation}
\int_{0}^{\infty}(D^{2}+\lambda)^{-n}\,d\lambda=\frac{1}{n-1}\,D^{2(1-n)},
\label{integrated}%
\end{equation}
we get that $Y$ is in $\Psi{\mathcal{D}}\cap\,OP^{-1}$ provided we control the
remainders. To control the remainder in (\ref{Rn}) one can use,
\begin{equation}
\int_{0}^{\infty}\Vert(X(D^{2}+\lambda)^{-1})^{3}\Vert\,d\lambda<\infty\,,
\label{size}%
\end{equation}
while the other terms are uniformly on $OP^{-N}$ since $D^{2}(D^{2}%
+\lambda)^{-1}$ is bounded by 1 in any ${\mathcal{H}}^{s}$.

To get (\ref{size}) since $X \in OP^{1}$ one can replace $X$ by $\vert D
\vert$ and only integrate from $\lambda= 1$ to $\infty$. Then the inequality
$D^{2} + \lambda\geq2 \, \vert D \vert\, \lambda^{1/2}$ gives the required
result. \endproof

\begin{lem}
\label{partial}

\begin{enumerate}
\item For any $N$ there is an element $B(t) \in\Psi{\mathcal{D}}$ such that
modulo $OP^{-N}$,
\begin{equation}
\frac{\partial}{\partial t} \, (\mathrm{Log} \, (D^{2} + tX) - \mathrm{Log}
D^{2} - \mathrm{Log} \, (1+tXD^{-2})) = [D^{2} + tX , B(t)] \label{eq1.33}%
\end{equation}

\item Modulo $OP^{-N}$ one has
\[
\mathrm{Log} \, (D^{2} + X) - \mathrm{Log} D^{2} - \mathrm{Log} \, (1+X
D^{-2}) = [D^{2} , B_{1}] + [X , B_{2}]
\]
where $B_{1} = \int_{0}^{1} B(t) \, dt \, , \ B_{2} = \int_{0}^{1} t \, B(t)
\, dt$ are in $\Psi{\mathcal{D}}$.
\end{enumerate}
\end{lem}

\proof1) From (\ref{logy}) one has,
\begin{equation}
\frac{\partial}{\partial t}\,\mathrm{Log}\,(D^{2}+tX)=\int_{0}^{\infty}%
\,\frac{1}{\lambda+D^{2}+tX}\ X\ \frac{1}{\lambda+D^{2}+tX}\,d\lambda,
\label{eq1.34}%
\end{equation}
while
\begin{equation}
\int_{0}^{\infty}X\,\frac{1}{(\lambda+D^{2}+tX)^{2}}\,d\lambda=X(D^{2}%
+tX)^{-1}, \label{integrated1}%
\end{equation}
which is the derivative in $t$ of $\mathrm{Log}\,(1+tXD^{-2})$ since,
$X(D^{2}+tX)^{-1}=XD^{-2}(1+tXD^{-2})^{-1}$.

We thus get, calling $Z(t)$ the left hand side of \eqref{eq1.33},
\begin{equation}
Z(t)=\int_{0}^{\infty}\left[  \frac{1}{\lambda+D^{2}+tX}\,,\ X\,\frac
{1}{\lambda+D^{2}+tX}\right]  \,d\lambda\,. \label{Zdet}%
\end{equation}
Let us define,
\begin{equation}
\nabla_{t}(T)=[D^{2}+tX,T]\,, \label{nablat}%
\end{equation}
and apply the formula of lemma~\ref{newton}~b) with $\lambda+D^{2}+tX$ instead
of $D^{2}$ and $T=X(\lambda+D^{2}+tX)^{-1}$. We thus get,
\begin{equation}
\left[  \frac{1}{\lambda+D^{2}+tX}\,,\ T\right]  =\sum_{1}^{n}\,(-1)^{k}%
\,\nabla_{t}^{k}(T(\lambda+D^{2}+tX)^{-(k+1)})+R_{n}, \label{nablat1}%
\end{equation}
where we put $(\lambda+D^{2}+tX)^{-(k+1)}$ inside the argument of $\nabla
_{t}^{k}$ since it is in the centralizer of $\nabla_{t}$. Thus,
\begin{equation}
\left[  \frac{1}{\lambda+D^{2}+tX}\,,\ T\right]  =\nabla_{t}\left(  \sum
_{1}^{n}\,(-1)^{k}\,\nabla_{t}^{k-1}(X)(\lambda+D^{2}+tX)^{-(k+2)}\right)
+R_{n}\,. \label{nablat2}%
\end{equation}
When integrated in $\lambda$ the parenthesis gives,
\begin{equation}
B(t)=\sum_{1}^{n}\,(-1)^{k}\,\nabla_{t}^{k-1}(X)\,\frac{1}{k+1}\ \frac
{1}{(D^{2}+tX)^{k+1}}\,. \label{nablat3}%
\end{equation}
Let us then check that $(D^{2}+tX)^{-1}\in\Psi{\mathcal{D}}$. We just expand
it as,
\begin{equation}
(D^{2}+tX)^{-1}=D^{-2}-D^{-2}\,t\,X\,D^{-2}+D^{-2}\,t\,X\,D^{-2}%
\,t\,X\,D^{-2}-\ldots\label{psidcheck}%
\end{equation}
It follows that $B(t)\in\Psi{\mathcal{D}}$ while,
\begin{equation}
Z(t)=\nabla_{t}(B(t))+R_{n}^{\prime}\,. \label{rprime}%
\end{equation}

2) Follows by integration using (\ref{nablat3}), (\ref{psidcheck}) to express
$B_{j}$ as explicit elements of $\Psi{\mathcal{D}}$ $\mathrm{mod} \, OP^{-N}$.
\endproof

\medskip

\subsection{The variation $\zeta_{D+A}(0)-\zeta_{D}(0)$}

\noindent\newline We are now ready to prove the main result of this section,
we work as above with a regular spectral triple with simple dimension spectrum.

\medskip

\begin{thm}
\label{theformula}

Let $A$ be a gauge potential,

\begin{enumerate}
\item The function $\zeta_{D+A}(s)$ extends to a meromorphic function with at
most simple poles.

\item It is regular at $s=0$.

\item One has
\[
\zeta_{D+A}(0)-\zeta_{D}(0)=\,-\,{\int\!\!\!\!\!\! -}\, \mathrm{Log}%
(1+\,A\,D^{-1})=\,\sum\,\frac{(-1)^{n}}{n}\,{\int\!\!\!\!\!\! -}\,
(A\,D^{-1})^{n}
\]

\end{enumerate}
\end{thm}

\proof1) We start from the expansional formula
\begin{equation}
e^{A+B}\,e^{-A}=\sum_{0}^{\infty}\int_{0\leq t_{1}\leq\cdots\leq t_{n}\leq
1}B(t_{1})\,B(t_{2})\ldots B(t_{n})\,\prod\,dt_{i} \label{expansional}%
\end{equation}
where
\begin{equation}
B(t)=e^{tA}\,B\,e^{-tA}\,.
\end{equation}
We take $A=-\frac{s}{2}\,\mathrm{Log}D^{2}$ and $B=-\frac{s}{2}\,Y$ so that,
\begin{equation}
e^{A+B}=(D^{2}+X)^{-s/2}\,,\ e^{A}=(D^{2})^{-s/2}\,.
\end{equation}
We define the one parameter group,
\begin{equation}
\sigma_{u}(T)=(D^{2})^{u/2}\,T\,(D^{2})^{-u/2}, \label{sigmau}%
\end{equation}
so that with the above notations we get,
\begin{equation}
B(t)=-\frac{s}{2}\ \sigma_{-st}(Y)\,. \label{bt}%
\end{equation}
We can thus write,
\begin{align}
&  (D^{2}+X)^{-s/2}=(D^{2})^{-s/2}+\sum_{n=1}^{\infty}\left(  -\frac{s}%
{2}\right)  ^{n}\label{expan}\\
&  \left(  \int_{0\leq t_{1}\leq\cdots\leq t_{n}\leq1}\sigma_{-st_{1}%
}(Y)\ldots\sigma_{-st_{i}}(Y)\ldots\sigma_{-st_{n}}(Y)\,\prod\,dt_{i}\right)
(D^{2})^{-s/2}\,.\nonumber
\end{align}
Since by lemma \ref{difflog} one has $Y\in\Psi{\mathcal{D}}\cap\,OP^{-1}$ for
any given half plane $H=\{z\,;\Re(z)\geq a\}$ only finitely many terms of the
sum \eqref{expan} contribute to the singularities in $H$ of the function
$\zeta_{D+A}(s)=\mathrm{Tr}((D^{2}+X)^{-s/2})$ and the expansion of the one
parameter group $\sigma_{u}$ (\textit{cf.\/}\ \cite{[C-M2]})
\begin{align}
\sigma_{2z}(T)  &  =T+z\,\epsilon(T)+\frac{z(z-1)}{2!}\,\epsilon^{2}%
(T)+\cdots\nonumber\\
&  +\frac{z(z-1)\cdots(z-n+1)}{n!}\,\epsilon^{n}(T)\qquad\mathrm{mod}%
\,OP^{q-(n+1)} \label{expsigma}%
\end{align}
where $T\in OP^{q}$ and,
\begin{equation}
\epsilon(T)=[D^{2},T]\,D^{-2}=[D^{2},TD^{-2}]\,. \label{epsilon}%
\end{equation}

gives the required meromorphic continuation.

2) By hypothesis the functions of the form $\mathrm{Tr}(P\,|D|^{-s})$ for
$P\in\Psi{\mathcal{D}}$ have at most simple poles thus only the first term of
the infinite sum in \eqref{expan} can contribute to the value $\zeta
_{D+A}(0)-\zeta_{D}(0)$. This first term is
\[
-\frac{s}{2}\int_{0}^{1}\,\sigma_{-st}(Y)\,dt\,(D^{2})^{-s/2},
\]
and using \eqref{expsigma} one can replace $\sigma_{-st}(Y)$ by $Y$ without
altering the value of $\zeta_{D+A}(0)-\zeta_{D}(0)$ which is hence, using the
definition of the residue
\begin{equation}
{\int\!\!\!\!\!\!-}\,P\,=\,\mathrm{Res}_{s=0}\,\mathrm{Tr}(P\,|D|^{-s}),
\label{resdef}%
\end{equation}
given by
\begin{equation}
\zeta_{D+A}(0)-\zeta_{D}(0)=\,-\frac{1}{2}\,{\int\!\!\!\!\!\!-}\;Y=\,-\frac
{1}{2}\,{\int\!\!\!\!\!\!-}\,\mathrm{Log}(1+\,X\,D^{-2}), \label{zetavar}%
\end{equation}
using Lemma \ref{partial} (2) and the trace property \eqref{residue}.

\medskip

3) For any elements $a,b\in\Psi{\mathcal{D}}\cap OP^{-1}$ one has the
identity
\begin{equation}
\,{\int\!\!\!\!\!\!-}\,\mathrm{Log}((1+a)(1+b))=\,{\int\!\!\!\!\!\!-}%
\,\mathrm{Log}(1+a)+\,{\int\!\!\!\!\!\!-}\,\mathrm{Log}(1+b). \label{tracelog}%
\end{equation}
This can be checked directly using the expansion
\[
\mathrm{Log}(1+a)=\,\sum_{1}^{\infty}\,(-1)^{n+1}\,\frac{a^{n}}{n},
\]
and the trace property \eqref{residue} of the residue. In fact one can reduce
it to the identity
\[
\,{\int\!\!\!\!\!\!-}\,(t+b)^{-1}\,(t+a)^{-1}\,(2t+a+b)=\,{\int\!\!\!\!\!\!-}%
\,((t+a)^{-1}+\,(t+b)^{-1}),
\]
which follows from \eqref{residue} and the equality
\[
(t+a)^{-1}\,(2t+a+b)\,(t+b)^{-1}=\,(t+a)^{-1}+\,(t+b)^{-1}.
\]
Applying \eqref{tracelog} to $a=D^{-1}\,A$ and $b=A\,D^{-1}$ one gets, with
$X=DA+AD+A^{2}$ as above,
\begin{equation}
{\int\!\!\!\!\!\!-}\,\mathrm{Log}(1+\,X\,D^{-2})=2\,{\int\!\!\!\!\!\!-}%
\,\mathrm{Log}(1+\,A\,D^{-1}), \label{tracelog1}%
\end{equation}
which combined with \eqref{zetavar} gives the required equality. \endproof

\bigskip

\section{Yang-Mills $+$ Chern-Simons}

\begin{figure}[ptb]
\begin{center}
\includegraphics[scale=0.5]{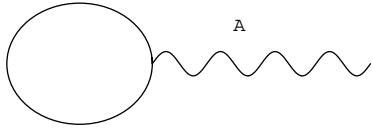}
\end{center}
\caption{The tadpole graph. }%
\label{Fig1leg}%
\end{figure}

We work in dimension $\leq4$ and make the following hypothesis of
\emph{vanishing tadpole} (\textit{cf.\/}\ Figure \ref{Fig1leg})
\begin{equation}
{\int\!\!\!\!\!\!-}\,a\,[D,b]\,D^{-1}=\,0{\,,\quad\forall\,}a\,,\;b\in
{\mathcal{A}}. \label{tadpole}%
\end{equation}
By Theorem \ref{theformula} this condition is equivalent to the vanishing of
the first order variation of the (scale independent part of) the spectral
action under inner fluctuations, and is thus a natural hypothesis.

Given a Hochschild cochain $\varphi$ of dimension $n$ on an algebra
${\mathcal{A}}$, normalized so that
\[
\varphi(a_{0},\,a_{1},\cdots,a_{n})=0,
\]
if any of the $a_{j}$ for $j>0$ is a scalar, it defines (\textit{cf.\/}%
\ \cite{book}) a functional on the universal $n$-forms $\Omega^{n}%
({\mathcal{A}})$ by the equality
\begin{equation}
\int_{\varphi}\,a_{0}\,da_{1}\cdots\,da_{n}=\;\varphi(a_{0},\,a_{1}%
,\cdots,a_{n}). \label{hoschtocyc}%
\end{equation}
When $\varphi$ is a Hochschild cocycle one has
\begin{equation}
\int_{\varphi}\,a\,\omega=\,\int_{\varphi}\,\omega\,a,\ \ \ \forall
a\in{\mathcal{A}}. \label{centr}%
\end{equation}
The boundary operator $B_{0}$ defined on normalized cochains by
\begin{equation}
(B_{0}\varphi)(a_{0},\,a_{1},\cdots,a_{n-1})=\,\varphi(1,a_{0},\,a_{1}%
,\cdots,a_{n-1}), \label{B0}%
\end{equation}
is defined in such a way that
\begin{equation}
\int_{\varphi}\,d\,\omega=\,\int_{B_{0}\varphi}\,\omega. \label{B0d}%
\end{equation}
Working in dimension $\leq4$ means that
\begin{equation}
D^{-1}\in{\mathcal{L}}^{(4,\infty)}, \label{dim4}%
\end{equation}
\textit{i.e.\/}\ that $D^{-1}$ is an infinitesimal of order $\frac{1}{4}$
(\textit{cf.\/}\ \cite{book}). The following functional is then a Hochschild
cocycle and is given as Dixmier trace of infinitesimals of order one,
\begin{equation}
\tau_{0}(a^{0},a^{1},a^{2},a^{3},a^{4})=\,{\int\!\!\!\!\!\!-}\,a^{0}%
\,[D,a^{1}]\,D^{-1}[D,a^{2}]\,D^{-1}[D,a^{3}]\,D^{-1}[D,a^{4}]\,D^{-1}.
\label{tau0}%
\end{equation}
The following functional uses the residue in an essential manner,
\begin{equation}
\varphi(a^{0},a^{1},a^{2},a^{3})=\ {\int\!\!\!\!\!\!-}a^{0}[D,a^{1}%
]\,D^{-1}[D,a^{2}]\,D^{-1}[D,a^{3}]\,D^{-1}. \label{phi}%
\end{equation}

\medskip

\begin{lem}
\label{basic}

\begin{enumerate}
\item $b \, \varphi= - \, \tau_{0}$

\item $b \, B_{0} \, \tau_{0} = 2 \, \tau_{0} $

\item $B_{0} \, \varphi= 0$
\end{enumerate}
\end{lem}

\proof1) One has,
\begin{align}
b\,\varphi(a^{0},\ldots,a^{4})  &  ={\int\!\!\!\!\!\!-}\,a^{0}\,a^{1}%
[D,a^{2}]\,D^{-1}\,[D,a^{3}]\,D^{-1}\,[D,a^{4}]\,D^{-1}\nonumber\\
&  -{\int\!\!\!\!\!\!-}\,a^{0}(a^{1}[D,a^{2}]+[D,a^{1}]\,a^{2})\,D^{-1}%
\,[D,a^{3}]\,D^{-1}\,[D,a^{4}]\,D^{-1}\nonumber\\
&  +{\int\!\!\!\!\!\!-}\,a^{0}[D,a^{1}]\,D^{-1}\,(a^{2}[D,a^{3}]+[D,a^{2}%
]\,a^{3})\,D^{-1}\,[D,a^{4}]\,D^{-1}\nonumber\\
&  -{\int\!\!\!\!\!\!-}\,a^{0}[D,a^{1}]\,D^{-1}\,[D,a^{2}]\,D^{-1}%
\,(a^{3}[D,a^{4}]+[D,a^{3}]\,a^{4})\,D^{-1}\nonumber\\
&  +{\int\!\!\!\!\!\!-}\,a^{4}\,a^{0}[D,a^{1}]\,D^{-1}\,[D,a^{2}%
]\,D^{-1}\,[D,a^{3}]\,D^{-1}\,.\nonumber
\end{align}
Thus using
\begin{equation}
a\,D^{-1}-D^{-1}\,a=D^{-1}\,[D,a]\,D^{-1}, \label{dinv}%
\end{equation}
we get 2).

2) One has
\[
B_{0}\,\tau_{0}(a^{0},a^{1},a^{2},a^{3})=\,{\int\!\!\!\!\!\!-}\,[D,a^{0}%
]\,D^{-1}[D,a^{1}]\,D^{-1}[D,a^{2}]\,D^{-1}[D,a^{3}]\,D^{-1}%
\]%
\[
=-\varphi(a^{0},a^{1},a^{2},a^{3})+\,\tilde{\varphi}(a^{0},a^{1},a^{2}%
,a^{3}),
\]
where
\[
\tilde{\varphi}(a^{0},a^{1},a^{2},a^{3})=\,{\int\!\!\!\!\!\!-}\,a^{0}%
\,D^{-1}[D,a^{1}]\,D^{-1}[D,a^{2}]\,D^{-1}[D,a^{3}].
\]
Thus it is enough to check that $b\,\tilde{\varphi}=\,\tau_{0}$. One has
\begin{align}
b\,\tilde{\varphi}(a^{0},\ldots,a^{4})  &  ={\int\!\!\!\!\!\!-}\,a^{0}%
\,a^{1}\,D^{-1}\,[D,a^{2}]\,D^{-1}\,[D,a^{3}]\,D^{-1}\,[D,a^{4}]\nonumber\\
&  -{\int\!\!\!\!\!\!-}\,a^{0}\,D^{-1}\,(a^{1}[D,a^{2}]+[D,a^{1}%
]\,a^{2})\,D^{-1}\,[D,a^{3}]\,D^{-1}\,[D,a^{4}]\nonumber\\
&  +{\int\!\!\!\!\!\!-}\,a^{0}\,D^{-1}\,[D,a^{1}]\,D^{-1}\,(a^{2}%
[D,a^{3}]+[D,a^{2}]\,a^{3})\,D^{-1}\,[D,a^{4}]\nonumber\\
&  -{\int\!\!\!\!\!\!-}\,a^{0}\,D^{-1}\,[D,a^{1}]\,D^{-1}\,[D,a^{2}%
]\,D^{-1}\,(a^{3}[D,a^{4}]+[D,a^{3}]\,a^{4})\nonumber\\
&  +{\int\!\!\!\!\!\!-}\,a^{4}\,a^{0}\,D^{-1}\,[D,a^{1}]\,D^{-1}%
\,[D,a^{2}]\,D^{-1}\,[D,a^{3}]\,\nonumber\\
&  ={\int\!\!\!\!\!\!-}\,a^{0}\,D^{-1}\,[D,a^{1}]\,D^{-1}[D,a^{2}%
]\,D^{-1}[D,a^{3}]\,D^{-1}[D,a^{4}]\,.\nonumber
\end{align}
and using \eqref{dinv} one gets the required equality since, using
\eqref{dim4},
\[
{\int\!\!\!\!\!\!-}\,a^{0}\,[D,a^{1}]\,D^{-1}[D,a^{2}]\,D^{-1}[D,a^{3}%
]\,D^{-1}[D,a^{4}]\,D^{-1}=\,{\int\!\!\!\!\!\!-}\,a^{0}\,D^{-1}\,[D,a^{1}%
]\,D^{-1}[D,a^{2}]\,D^{-1}[D,a^{3}]\,D^{-1}[D,a^{4}].
\]

3) We use the notation
\begin{equation}
\alpha(a)=D\,a\,D^{-1}{\,,\quad\forall\,}a\in{\mathcal{A}}.
\label{definealpha}%
\end{equation}
Note that in general $\alpha(a)\notin{\mathcal{A}}$. One has
\[
\alpha(ab)=\alpha(a)\,\alpha(b){\,,\quad\forall\,}a\,,\;b\in{\mathcal{A}}.
\]
Let us show that the tadpole hypothesis \eqref{tadpole} implies that for any
three elements $a,b,c\in{\mathcal{A}}$,
\begin{equation}
{\int\!\!\!\!\!\!-}\,\alpha^{\epsilon_{1}}(a)\,\alpha^{\epsilon_{2}%
}(b)\,\alpha^{\epsilon_{3}}(c)=\,{\int\!\!\!\!\!\!-}\,a\,b\,c,
\label{removealpha}%
\end{equation}
for all $\epsilon_{j}\in\{0,1\}$. The trace property of the residue shows that
this holds when all $\epsilon_{j}=1$. One is thus reduced to show that
\[
{\int\!\!\!\!\!\!-}\,\alpha(x)\,y=\,{\int\!\!\!\!\!\!-}\,x\,y{\,,\quad
\forall\,}x\,,\;y\in{\mathcal{A}},
\]
which follows from \eqref{tadpole}. One has by construction
\[
B_{0}\,\varphi(a_{0},a_{1},a_{2})=\,{\int\!\!\!\!\!\!-}\,(\alpha(a_{0}%
)-a_{0})(\alpha(a_{1})-a_{1})(\alpha(a_{2})-a_{2}),
\]
which vanishes since the terms cancel pairwise. \endproof

\medskip

\begin{lem}
\label{2phi} One has for any $A\in\Omega^{1}$ the equality
\[
{\int\!\!\!\!\!\!-}\,A\,D^{-1}\,A\,D^{-1}=\,-\,\int_{\varphi}\,A\,dA.
\]

\end{lem}

\proof Let us first show that for any $a_{j}\in{\mathcal{A}}$ one has
\begin{equation}
{\int\!\!\!\!\!\!-}a_{0}[D,a_{1}]\,D^{-1}\,a_{2}[D,a_{3}]\,D^{-1}%
=\,-\varphi(a^{0},a^{1},a^{2},a^{3}). \label{pol2}%
\end{equation}
It suffices using \eqref{dinv} to show that
\[
{\int\!\!\!\!\!\!-}a_{0}[D,a_{1}]\,a_{2}\,D^{-1}\,[D,a_{3}]\,D^{-1}=\,0,
\]
which follows using
\[
a_{0}\,[D,a_{1}]\,a_{2}=\,a_{0}\,[D,a_{1}\,a_{2}]-\,a_{0}\,a_{1}\,[D,a_{2}],
\]
and the vanishing of
\[
{\int\!\!\!\!\!\!-}a\,[D,b]\,D^{-1}\,[D,c]\,D^{-1}=\,{\int\!\!\!\!\!\!-}%
\,a\,(\alpha(b)-b)\,(\alpha(c)-c)=\,0, \quad\forall\, a,b,c\in{\mathcal{A}}
\]
using \eqref{tau0}. Let then $A_{1}=a_{0}\,da_{1}$, $A_{2}=a_{2}\,da_{3}$, one
has
\[
{\int\!\!\!\!\!\!-}\,A_{1}\,D^{-1}\,A_{2}\,D^{-1}=\,-\,\int_{\varphi}%
\,A_{1}\,dA_{2},
\]
since $dA_{2}=\,da_{2}\,da_{3}$, and the same holds for any $A_{j}\in
\Omega^{1}$ so that lemma \ref{2phi} follows. \endproof

\begin{lem}
\label{4phi} One has for any $A \in\Omega^{1}$ the equality
\[
{\int\!\!\!\!\!\! -}\,(A\, D^{-1})^{4}=\,\int_{\tau_{0}}\,A^{4}
\]

\end{lem}

\proof It is enough to check that with $a_{j}$, $b_{j}$ in ${\mathcal{A}}$ one
has
\[
\int_{\tau_{0}}\,a_{1}\,db_{1}\,a_{2}\,db_{2}\,a_{3}\,db_{3}\,a_{4}%
\,db_{4}=\,{\int\!\!\!\!\!\!-}\,A_{1}\,D^{-1}\,A_{2}\,D^{-1}\,A_{3}%
\,D^{-1}\,A_{4}\,D^{-1}\,,\quad A_{j}=\,a_{j}\,[D,b_{j}].
\]
Since there are $4$ terms $D^{-1}$ one is in the domain of the Dixmier trace
and one can freely permute the factors $\,D^{-1}$ with the elements of
${\mathcal{A}}$ in computing the residue of the right hand side. One can thus
assume that $a_{2}=a_{3}=a_{4}=1$. The result then follows from \eqref{tau0}.
\endproof

\begin{lem}
\label{phi3} One has for any $A_{j} \in\Omega^{1}$ the equality
\begin{align}
&  {\int\!\!\!\!\!\! -} A_{1} \, D^{-1} \, A_{2} \, D^{-1} \, A_{3}
\,D^{-1}=\,\int_{\varphi+ \frac{1}{2} \, B_{0} \, \tau_{0}} \, A_{1} \, A_{2}
\, A_{3}\\
&  - \, \frac{1}{2} \left(  \int_{\tau_{0}} (d A_{1}) \, A_{2} \, A_{3} +
\int_{\tau_{0}} \, A_{1} \, dA_{2} \, A_{3} + \int_{\tau_{0}} \, A_{1} \,
A_{2} \, d A_{3} \right)  \, .\nonumber
\end{align}

\end{lem}

\proof We can take $A_{j}=a_{j}\,db_{j}$ and the first task is to reorder
\begin{align}
a_{1}\,db_{1}\,a_{2}\,db_{2}\,a_{3}\,db_{3}  &  =a_{1}\,db_{1}\,a_{2}%
(d(b_{2}\,a_{3})-b_{2}\,da_{3})\,db_{3}\label{reorder}\\
&  =a_{1}(d(b_{1}\,a_{2})-b_{1}\,da_{2})\,d(b_{2}\,a_{3})\,db_{3}\nonumber\\
&  -a_{1}(d(b_{1}\,a_{2}\,b_{2})-b_{1}\,d(a_{2}\,b_{2}))\,da_{3}%
\,db_{3}\nonumber\\
&  =a_{1}\,d(b_{1}\,a_{2})\,d(b_{2}\,a_{3})\,db_{3}-a_{1}\,b_{1}%
\,da_{2}\,d(b_{2}\,a_{3})\,db_{3}\nonumber\\
&  -a_{1}\,d(b_{1}\,a_{2}\,b_{2})\,da_{3}\,db_{3}+a_{1}\,b_{1}\,d(a_{2}%
\,b_{2})\,da_{3}\,db_{3}\,.\nonumber
\end{align}
We thus get
\begin{align}
\int_{\varphi}\,A_{1}\,A_{2}\,A_{3}  &  ={\int\!\!\!\!\!\!-}a_{1}%
[D,b_{1}\,a_{2}]\,D^{-1}\,[D,b_{2}\,a_{3}]\,D^{-1}\,[D,b_{3}]\,D^{-1}%
\nonumber\\
&  -{\int\!\!\!\!\!\!-}a_{1}\,b_{1}[D,a_{2}]\,D^{-1}\,[D,b_{2}\,a_{3}%
]\,D^{-1}\,[D,b_{3}]\,D^{-1}\nonumber\\
&  -{\int\!\!\!\!\!\!-}a_{1}[D,b_{1}\,a_{2}\,b_{2}]\,D^{-1}\,[D,a_{3}%
]\,D^{-1}\,[D,b_{3}]\,D^{-1}\nonumber\\
&  +{\int\!\!\!\!\!\!-}a_{1}\,b_{1}[D,a_{2}\,b_{2}]\,D^{-1}\,[D,a_{3}%
]\,D^{-1}\,[D,b_{3}]\,D^{-1}\nonumber\\
&  ={\int\!\!\!\!\!\!-}a_{1}[D,b_{1}]\,a_{2}\,D^{-1}\,[D,b_{2}\,a_{3}%
]\,D^{-1}\,[D,b_{3}]\,D^{-1}\nonumber\\
&  -{\int\!\!\!\!\!\!-}a_{1}[D,b_{1}]\,a_{2}\,b_{2}\,D^{-1}\,[D,a_{3}%
]\,D^{-1}\,[D,b_{3}]\,D^{-1}\,.\nonumber
\end{align}
Using $[D^{-1},b_{2}]=-D^{-1}\,[D,b_{2}]\,D^{-1}$ we thus get,
\begin{align}
\int_{\varphi}\,A_{1}\,A_{2}\,A_{3}  &  ={\int\!\!\!\!\!\!-}a_{1}%
[D,b_{1}]\,a_{2}\,D^{-1}\,[D,b_{2}]\,a_{3}\,D^{-1}\,[D,b_{3}]\,D^{-1}%
\label{phiphi}\\
&  -{\int\!\!\!\!\!\!-}a_{1}[D,b_{1}]\,a_{2}\,D^{-1}\,[D,b_{2}]\,D^{-1}%
\,[D,a_{3}]\,D^{-1}\,[D,b_{3}]\,D^{-1}\,.\nonumber
\end{align}
Next one has using \eqref{reorder}
\begin{align}
\int_{B_{0}\tau_{0}}\,A_{1}\,A_{2}\,A_{3}  &  ={\int\!\!\!\!\!\!-}%
[D,a_{1}]\,D^{-1}\,[D,b_{1}\,a_{2}]\,D^{-1}\,[D,b_{2}\,a_{3}]\,D^{-1}%
\,[D,b_{3}]\,D^{-1}\nonumber\\
&  -{\int\!\!\!\!\!\!-}[D,a_{1}\,b_{1}]\,D^{-1}\,[D,a_{2}]\,D^{-1}%
\,[D,b_{2}\,a_{3}]\,D^{-1}\,[D,b_{3}]\,D^{-1}\nonumber\\
&  -{\int\!\!\!\!\!\!-}[D,a_{1}]\,D^{-1}\,[D,b_{1}\,a_{2}\,b_{2}%
]\,D^{-1}\,[D,a_{3}]\,D^{-1}\,[D,b_{3}]\,D^{-1}\nonumber\\
&  +{\int\!\!\!\!\!\!-}[D,a_{1}\,b_{1}]\,D^{-1}\,[D,a_{2}\,b_{2}%
]\,D^{-1}\,[D,a_{3}]\,D^{-1}\,[D,b_{3}]\,D^{-1}\,.\nonumber
\end{align}
Since one is in the domain of the Dixmier trace, one can permute $D^{-1}$ with
$a$ for $a\in{\mathcal{A}}$. Thus the first two terms combine to give,
\begin{align}
&  {\int\!\!\!\!\!\!-}[D,a_{1}]\,D^{-1}\,[D,b_{1}]\,a_{2}\,D^{-1}%
\,[D,b_{2}\,a_{3}]\,D^{-1}\,[D,b_{3}]\,D^{-1}\nonumber\\
&  -{\int\!\!\!\!\!\!-}a_{1}[D,b_{1}]\,D^{-1}\,[D,a_{2}]\,D^{-1}%
\,[D,b_{2}\,a_{3}]\,D^{-1}\,[D,b_{3}]\,D^{-1},\nonumber
\end{align}
and the last two terms combine to give,
\begin{align}
&  {\int\!\!\!\!\!\!-}a_{1}[D,b_{1}]\,D^{-1}\,[D,a_{2}\,b_{2}]\,D^{-1}%
\,[D,a_{3}]\,D^{-1}\,[D,b_{3}]\,D^{-1}\nonumber\\
&  -{\int\!\!\!\!\!\!-}[D,a_{1}]\,D^{-1}\,[D,b_{1}]\,a_{2}\,b_{2}%
\,D^{-1}\,[D,a_{3}]\,D^{-1}\,[D,b_{3}]\,D^{-1}\,.\nonumber
\end{align}
Thus these 4 terms add up to give
\begin{align}
\int_{B_{0}\tau_{0}}\,A_{1}\,A_{2}\,A_{3}  &  ={\int\!\!\!\!\!\!-}%
[D,a_{1}]\,D^{-1}\,[D,b_{1}]\,a_{2}\,D^{-1}\,[D,b_{2}]\,a_{3}\,D^{-1}%
\,[D,b_{3}]\,D^{-1}\label{boto}\\
&  -{\int\!\!\!\!\!\!-}a_{1}[D,b_{1}]\,D^{-1}\,[D,a_{2}]\,D^{-1}%
\,[D,b_{2}]\,a_{3}\,D^{-1}\,[D,b_{3}]\,D^{-1}\nonumber\\
&  +{\int\!\!\!\!\!\!-}a_{1}[D,b_{1}]\,D^{-1}\,a_{2}[D,b_{2}]\,D^{-1}%
\,[D,a_{3}]\,D^{-1}\,[D,b_{3}]\,D^{-1}\,.\nonumber
\end{align}
Combining this with \eqref{phiphi} thus gives,
\begin{align}
\int_{\varphi+\frac{1}{2}\,B_{0}\tau_{0}}\,A_{1}\,A_{2}\,A_{3}  &
={\int\!\!\!\!\!\!-}\,a_{1}\,[D,b_{1}]\,a_{2}\,D^{-1}\,[D,b_{2}]\,a_{3}%
\,D^{-1}\,[D,b_{3}]\,D^{-1}\label{full}\\
&  +\frac{1}{2}\ {\int\!\!\!\!\!\!-}[D,a_{1}]\,D^{-1}\,[D,b_{1}]\,D^{-1}%
\,a_{2}[D,b_{2}]\,D^{-1}\,a_{3}[D,b_{3}]\,D^{-1}\nonumber\\
&  -\frac{1}{2}\ {\int\!\!\!\!\!\!-}a_{1}[D,b_{1}]\,D^{-1}\,[D,a_{2}%
]\,D^{-1}\,[D,b_{2}]\,D^{-1}\,a_{3}[D,b_{3}]\,D^{-1}\nonumber\\
&  -\frac{1}{2}\ {\int\!\!\!\!\!\!-}a_{1}[D,b_{1}]\,D^{-1}\,a_{2}%
\,[D,b_{2}]\,D^{-1}\,[D,a_{3}]\,D^{-1}\,[D,b_{3}]\,D^{-1}\,.\nonumber
\end{align}
But one has, using $[a,D^{-1}]=D^{-1}\,[D,a]\,D^{-1}$,
\begin{align}
&  {\int\!\!\!\!\!\!-}a_{1}[D,b_{1}]\,a_{2}\,D^{-1}\,[D,b_{2}]\,a_{3}%
\,D^{-1}\,[D,b_{3}]\,D^{-1}\nonumber\\
&  ={\int\!\!\!\!\!\!-}a_{1}[D,b_{1}]\,D^{-1}\,[D,a_{2}]\,D^{-1}%
\,[D,b_{2}]\,a_{3}\,D^{-1}\,[D,b_{3}]\,D^{-1}\nonumber\\
&  +{\int\!\!\!\!\!\!-}a_{1}[D,b_{1}]\,D^{-1}\,a_{2}[D,b_{2}]\,a_{3}%
\,D^{-1}\,[D,b_{3}]\,D^{-1}\nonumber\\
&  ={\int\!\!\!\!\!\!-}a_{1}[D,b_{1}]\,D^{-1}\,[D,a_{2}]\,D^{-1}%
\,[D,b_{2}]\,D^{-1}\,a_{3}[D,b_{3}]\,D^{-1}\nonumber\\
&  +{\int\!\!\!\!\!\!-}a_{1}[D,b_{1}]\,D^{-1}\,a_{2}[D,b_{2}]\,D^{-1}%
\,[D,a_{3}]\,D^{-1}\,[D,b_{3}]\,D^{-1}\nonumber\\
&  +{\int\!\!\!\!\!\!-}A_{1}\,D^{-1}\,A_{2}\,D^{-1}\,A_{3}\,D^{-1},\nonumber
\end{align}
which combined with \eqref{full} gives the required equality. \endproof

We can now state the main result

\begin{thm}
\label{main} Under the tadpole hypothesis \eqref{tadpole} one has

\begin{enumerate}
\item $\psi= \,\varphi+\frac12 B_{0} \tau_{0}$ is a cyclic $3$-cocycle given
(with $\alpha(x)=D\,x\,D^{-1}$) by
\begin{equation}
\label{simplepsi}\psi(a_{0},a_{1},a_{2},a_{3})=\,\frac12\,{\int\!\!\!\!\!\!
-}\,(\alpha(a_{0})\,a_{1}\,\alpha(a_{2})\,a_{3}-\,a_{0}\,\alpha(a_{1}%
)\,a_{2}\,\alpha(a_{3}))
\end{equation}

\item For any $A \in\Omega^{1}$ one has
\begin{equation}
\label{fundthm}\,{\int\!\!\!\!\!\! -} \mathrm{Log} (1 + AD^{-1}) =\,-\frac14
\,\int_{\tau_{0}}\,(dA+\,A^{2})^{2}+ \frac12\,\int_{\psi}\,(AdA+\,\frac23
A^{3})
\end{equation}

\end{enumerate}
\end{thm}

\proof1) By lemma \ref{basic} $\psi$ is a Hochschild cocycle. Moreover by
lemma \ref{basic} it is in the kernel of $B_{0}$ and is hence cyclic.
Expanding the expression
\[
\psi(a_{0},a_{1},a_{2},a_{3})=\,\frac{1}{2}\,{\int\!\!\!\!\!\!-}%
\,(\alpha(a_{0})+\,a_{0})\,(\alpha(a_{1})-\,a_{1})\,(\alpha(a_{2}%
)-\,a_{2})\,(\alpha(a_{3})-\,a_{3}),
\]
and using \eqref{removealpha}, one gets \eqref{simplepsi}.

2) One has
\[
{\int\!\!\!\!\!\!-}\mathrm{Log}(1+AD^{-1})=\,-\frac{1}{2}\ {\int
\!\!\!\!\!\!-}(AD^{-1})^{2}+\frac{1}{3}\ {\int\!\!\!\!\!\!-}(AD^{-1}%
)^{3}-\frac{1}{4}\ {\int\!\!\!\!\!\!-}\,(AD^{-1})^{4}.
\]
Both sides of \eqref{fundthm} are thus polynomials in $A$ and it is enough to
compare the monomials of degree $2$, $3$ and $4$. In degree $2$ the right hand
side of \eqref{fundthm} gives
\[
-\frac{1}{4}\,\int_{\tau_{0}}\,(dA)^{2}+\frac{1}{2}\,\int_{\psi}\,AdA=\frac
{1}{2}\,\int_{\varphi}\,AdA,
\]
using \eqref{B0d}. Thus by lemma \ref{2phi} one gets the same as the term of
degree two in the left hand side of \eqref{fundthm}. In degree $4$ the right
hand side of \eqref{fundthm} gives
\[
-\frac{1}{4}\,\int_{\tau_{0}}\,A^{4}=\,-\frac{1}{4}\ {\int\!\!\!\!\!\!-}%
\,(AD^{-1})^{4},
\]
by lemma \ref{4phi}. It remains to handle the cubic terms, the right hand side
of \eqref{fundthm} gives
\[
-\frac{1}{4}\,\int_{\tau_{0}}\,(dA\,A^{2}+A^{2}\,dA)+\frac{1}{3}\,\int_{\psi
}\,A^{3},
\]
which using lemma \ref{phi3} gives
\[
\frac{1}{3}\,{\int\!\!\!\!\!\!-}\,(A\,D^{-1})^{3}+\frac{1}{6}\int_{\tau_{0}%
}(dA\,A^{2}+A\,dA\,A+\,A^{2}\,dA)-\frac{1}{4}\,\int_{\tau_{0}}\,(dA\,A^{2}%
+A^{2}\,dA).
\]
Thus it remains to show that the sum of the last two terms is zero. In fact
\[
\int_{\tau_{0}}dA\,A^{2}=\,\int_{\tau_{0}}A\,dA\,A=\,\int_{\tau_{0}}%
\,A^{2}\,dA.
\]
This follows from the more general equality
\begin{equation}
\int_{\tau_{0}}\,\omega_{1}\,\omega_{2}\,\omega_{3}\,\omega_{4}\,=\,\int
_{\tau_{0}}\,\omega_{2}\,\omega_{3}\,\omega_{4}\,\omega_{1}{\,,\quad\forall
\,}\,\omega_{j}\in\Omega^{1}, \label{cyclictr}%
\end{equation}
which is seen as follows. Let $\omega_{j}=a_{j}\,db_{j}$, then
\[
\int_{\tau_{0}}\,\omega_{1}\,\omega_{2}\,\omega_{3}\,\omega_{4}\,=\,{\int
\!\!\!\!\!\!-}\,a_{1}[D,b_{1}]\,D^{-1}\,a_{2}[D,b_{2}]\,D^{-1}\,a_{3}%
[D,b_{3}]\,D^{-1}\,a_{4}[D,b_{4}]\,D^{-1},
\]
so that \eqref{cyclictr} follows from the trace property of the residue.
\endproof

\medskip Combining this result with Theorem \ref{theformula} one gets \medskip

\begin{cor}
The variation under inner fluctuations of the scale independent terms of the
spectral action is given in dimension $4$ by
\[
\zeta_{D+A}(0)-\zeta_{D}(0)=\, \,\frac14 \,\int_{\tau_{0}}\,(dA+\,A^{2})^{2}-
\frac12\,\int_{\psi}\,(AdA+\,\frac23 A^{3})
\]

\end{cor}

\medskip

Note that there is still some freedom in the choice of the cocycles $\tau_{0}$
and $\psi$ involved in Theorem \ref{main},. Indeed let $B=A\,B_{0}$ be the
fundamental boundary operator in cyclic cohomology (\cite{book}), one has

\begin{prop}
\label{ambig} 1) Theorem \eqref{main} still holds after the replacements
$\tau_{0}\to\tau_{0} +\rho$ and $\psi\to\psi+\frac12 B_{0}\rho$ for any
Hochschild $4$-cocycle $\rho$ such that $B_{0} \, \rho$ is already cyclic
\textit{i.e.\/}\ such that $A \, B_{0} \, \rho= 4 \, B_{0} \, \rho$.

2) If the cocycle $\psi$ is in the image of $B$ \textit{i.e.\/}\ if $\psi\in
B(Z^{4}({\mathcal{A}},{\mathcal{A}}^{*}))$ one can eliminate $\psi$ by a
redefinition of $\tau_{0}$.
\end{prop}

\proof1) We first show that $\int_{\rho}$ is a graded trace (\textit{cf.\/}%
\ \cite{book}, Chapter III lemma 18). First since $\rho$ is a Hochschild
cocycle one has
\[
\int_{\rho}\,a\,\omega=\,\int_{\rho}\,\omega\,a{\,,\quad\forall\,}%
a\in{\mathcal{A}}.
\]
To show that $\int_{\rho}$ is a graded trace it is enough to check that
\[
\int_{\rho}\,da\,(a_{0}\,da_{1}\,da_{2}\,da_{3})=\,-\,\int_{\rho}%
\,a_{0}\,da_{1}\,da_{2}\,da_{3}\,da,
\]
\textit{i.e.\/}\ that
\[
B_{0}\,\rho(a\,a_{0},\ldots,a_{3})-\rho(a,a_{0},\ldots,a_{3})=\,-\,\rho
(a_{0},\ldots,a_{3},a),
\]
which follows (\textit{cf.\/}\ \cite{book}, Chapter III lemma 18) from
\[
B_{0}\,b+\,b^{\prime}\,B_{0}=\mathrm{id}-\lambda,
\]
(where $\lambda$ is the cyclic permutation) and $b\,\rho=0$, $b\,B_{0}%
\,\rho=0$. We need to show that the right hand side of \eqref{fundthm} is
unaltered by the above replacements. For the terms of degree $4$ one has to
show that%
\[
\int_{\rho}\,A^{4}=0,
\]
which holds because $\int_{\rho}$ is a graded trace. For the terms of degree
$3$ one has
\[
\int_{\rho}\,(dA\,A^{2}+\,A^{2}\,dA)-\frac{4}{3}\,\int_{\frac{1}{2}B_{0}%
\,\rho}\,A^{3}=\,\int_{\rho}\,(dA\,A^{2}+\,A^{2}\,dA-\frac{2}{3}\,d(A^{3})),
\]
and the graded trace property of $\int_{\rho}$ shows that this vanishes. For
the quadratic terms one has
\[
\int_{\rho}\,(dA)^{2}-2\,\int_{\frac{1}{2}B_{0}\,\rho}\,AdA=\,\int_{\rho
}\,((dA)^{2}-d(AdA))=0.
\]

2) By \cite{book} Chapter III, Lemma 19, the condition $\psi\in B(Z^{4}%
({\mathcal{A}},{\mathcal{A}}^{*}))$ implies that one can find a Hochschild
$4$-cocycle $\rho$ such that $\,B_{0} \, \rho$ is already cyclic and equal to
$-2\,\psi$ thus using 1) one can eliminate $\psi$. \endproof

The above ambiguity can thus be written in the form
\begin{equation}
\label{ambiguity}\psi\to\psi+ \delta{\,,\quad\forall\,} \delta\in
B(Z^{4}({\mathcal{A}},{\mathcal{A}}^{*}))\,.
\end{equation}
and it does not alter the periodic cyclic cohomology class of the three
cocycle $\psi$.

\medskip

The Yang-Mills action given by
\[
YM_{\tau}(A)=\,\int_{\tau}\,(dA+A^{2})^{2},
\]
is automatically gauge invariant under the gauge transformations
\begin{equation}
A\rightarrow\gamma_{u}(A)=\,u\,du^{\ast}\,+u\,A\,u^{\ast}{\,,\quad\forall
\,}u\in{\mathcal{A}}\,,\quad u\,u^{\ast}=u^{\ast}\,u=1, \label{gauge}%
\end{equation}
as soon as $\tau$ is a Hochschild cocycle since $F(A)=dA+A^{2}$ transforms
covariantly \textit{i.e.\/}\ $F(\gamma_{u}(A))=\,u\,F(A)\,u^{\ast}$. This
action and its precise relation with the usual Yang-Mills functional is
discussed at length in \cite{book} Chapter VI.

We now discuss briefly the invariance of the Chern-Simons action. An early
instance of this action in terms of cyclic cohomology can be found in
\cite{witten}. It is not in general invariant under gauge transformations but
one has the following more subtle invariance, \medskip

\begin{prop}
\label{sub} Let $\psi$ be a cyclic three cocycle on ${\mathcal{A}}$. The
functional
\[
CS_{\psi} (A) = \int_{\psi} \, A \, d \, A + \frac{2}{3} \, A^{3}
\]
fulfills the following invariance rule under the gauge transformation
$\gamma_{u} (A)=\,u\,du^{*}\,+u\,A\,u^{*}$,
\[
CS_{\psi} (\gamma_{u} (A)) = CS_{\psi} (A) + \frac{1}{3} \, \langle\psi, u
\rangle
\]
where $\langle\psi, u \rangle$ is the pairing between $HC^{3} ({\mathcal{A}})$
and $K_{1} ({\mathcal{A}})$.
\end{prop}

\proof let $A^{\prime}=\gamma_{u}(A)=\,u\,du^{\ast}\,+u\,A\,u^{\ast}$. One
has
\[
dA^{\prime}=\,du\,du^{\ast}\,+\,du\,A\,u^{\ast}+\,u\,dA\,u^{\ast
}-\,u\,A\,du^{\ast},
\]%
\[
A^{\prime}\,dA^{\prime}=\,u\,du^{\ast}\,du\,du^{\ast}+\,u\,du^{\ast
}\,du\,A\,u^{\ast}+\,u\,du^{\ast}\,u\,dA\,u^{\ast}-\,u\,du^{\ast
}\,u\,A\,du^{\ast}%
\]%
\[
+u\,A\,u^{\ast}\,du\,du^{\ast}+u\,A\,u^{\ast}\,du\,A\,u^{\ast}%
+u\,A\,dA\,u^{\ast}-u\,A^{2}\,du^{\ast}.
\]
So that using the graded trace property of $\int_{\psi}$ one gets
\[
\int_{\psi}(\,A^{\prime}\,dA^{\prime}-\,A\,dA)=\,
\]%
\[
\int_{\psi}\,(u\,du^{\ast}\,du\,du^{\ast}+\,du^{\ast}\,du\,A-\,du^{\ast
}\,u\,du^{\ast}\,u\,A+\,u^{\ast}\,du\,du^{\ast}\,u\,A+\,du^{\ast
}\,u\,dA+\,u^{\ast}\,du\,A^{2}-\,du^{\ast}\,u\,A^{2}),
\]
which using
\[
\int_{\psi}\,du^{\ast}\,u\,dA=\,-\,\int_{\psi}\,du^{\ast}\,du\,A,
\]
gives
\[
\int_{\psi}(\,A^{\prime}\,dA^{\prime}-\,A\,dA)=\,\int_{\psi}\,(u\,du^{\ast
}\,du\,du^{\ast}+2\,u^{\ast}\,du\,du^{\ast}\,u\,A+2\,u^{\ast}\,du\,A^{2}).
\]
Next one has
\[
\int_{\psi}\,(A^{^{\prime}\,3}-A^{3})=\,\int_{\psi}\,((u\,du^{\ast}%
)^{3}+\,3\,(u\,du^{\ast})^{2}\,u\,A\,u^{\ast}+\,3\,u\,du^{\ast}\,u\,A^{2}%
\,u^{\ast}).
\]
Since $du^{\ast}\,u=\,-\,u^{\ast}\,du$, the terms in $A^{2}$ cancel in the
variation of $CS_{\psi}$. Similarly one has $du^{\ast}\,u\,du^{\ast
}\,u=-\,u^{\ast}\,du\,du^{\ast}\,u$ so that the terms in $A$ also cancel. One
thus obtains
\[
CS_{\psi}(\gamma_{u}(A))-CS_{\psi}(A)=\,\int_{\psi}\,(u\,du^{\ast
}\,du\,du^{\ast}+\frac{2}{3}(u\,du^{\ast})^{3}).
\]
One has $(u\,du^{\ast})^{3}=-u\,du^{\ast}\,du\,du^{\ast}$ which gives the
required result. \endproof

\medskip

\begin{cor}
Let $\psi$ be the cyclic three cocycle of Theorem \ref{main} then its pairing
with the $K_{1}$-group vanishes identically,
\[
\langle\psi, u \rangle=0 {\,,\quad\forall\,} u\in K_{1} ({\mathcal{A}})
\]

\end{cor}

\proof The effect of the gauge transformation \eqref{gauge} is to replace the
operator $D+A$ by the unitarily equivalent operator $D+\gamma_{u}%
(A)=u(D+A)u^{\ast}$, thus the spectral invariants are unaltered by such a
transformation. Since the Yang-Mills term
\[
\frac{1}{4}\,\int_{\tau_{0}}\,(dA+\,A^{2})^{2},
\]
is invariant under gauge transformations, it follows that so is the
Chern-Simons term which implies by Proposition \ref{sub} that the pairing
between the cyclic cocycle $\psi$ and the unitary $u$ is zero. Tensoring the
original spectral triple by the finite geometry $(M_{n}({\mathbb{C}%
}),{\mathbb{C}}^{n},0)$ allows to apply the same argument to unitaries in
$M_{n}({\mathcal{A}})$ and shows that the pairing with the $K_{1}$-group
vanishes identically. \endproof

\bigskip

\section{Open questions}

We shall briefly discuss two important questions which are left open in the
generality of the above framework.

\subsection{Triviality of $\psi$}

\noindent\newline It is true under mild hypothesis that the vanishing of the
pairing with the $K_{1}$-group
\[
\langle\psi,u\rangle=0{\,,\quad\forall\,}u\in K_{1}({\mathcal{A}}),
\]
implies that the cyclic cocycle $\psi$ is homologous to zero,
\[
\psi\in B\,Z^{4}({\mathcal{A}},{\mathcal{A}}^{\ast}).
\]
Thus one can in any such case eliminate the Chern-Simons term using
Proposition \ref{ambig} 2). We have not been able to find an example where
$\psi$ does not belong to the image of $B$ and it could thus be that $\psi\in
B\,Z^{4}({\mathcal{A}},{\mathcal{A}}^{\ast})$ holds in full generality.

\subsection{Positivity}

\noindent\newline In a similar manner the freedom given by Proposition
\ref{ambig} should be used to replace the Hochschild cocycle $\tau_{0}$ by a
\emph{positive} Hochschild cocycle $\tau$. Positivity in Hochschild cohomology
was defined in \cite{concun} as the condition
\[
\int_{\tau}\,\omega\,\omega^{\ast}\geq0{\,,\quad\forall\,}\omega\in\Omega
^{2},
\]
where the adjoint $\omega^{\ast}$ is defined by
\[
(a_{0}\,da_{1}\,da_{2})^{\ast}=\,da_{2}^{\ast}\,da_{1}^{\ast}\,a_{0}^{\ast
}{\,,\quad\forall\,}a_{j}\in{\mathcal{A}}.
\]
It then follows easily (\textit{cf.\/}\ \cite{book} Chapter VI) that the
Yang-Mills action functional fulfills
\[
YM_{\tau}(A)\geq0{\,,\quad\forall\,}A\in\Omega^{1}\,.
\]

\section{Acknowledgment}

The research of A. Chamseddine is supported in part by the National Science
Foundation under Grant No. Phys-0313416.

\end{document}